\documentclass[aps,prd,twocolumn,groupedaddress,nofootinbib]{revtex4}
\pdfoutput=1  
\usepackage{bm}
\usepackage{graphicx,color}
\usepackage{latexsym,amsmath,amssymb,graphicx,booktabs}
\usepackage{hyperref}
\usepackage{placeins}
\usepackage{bm}
\usepackage{subfigure}

\definecolor{MyBlue}{rgb}{0.15,0.15,0.70}
\definecolor{Dgreen}{rgb}{0,0.7,0.0}

\hypersetup{
colorlinks=true,
citecolor=MyBlue,
linkcolor=MyBlue,
urlcolor=MyBlue
}

\usepackage{amssymb}
\usepackage{amsmath}
\usepackage{amsfonts}
\usepackage{upgreek}
\usepackage{latexsym}
\usepackage{appendix}
\usepackage{eufrak}
\usepackage{dsfont}

\usepackage[export]{adjustbox}

\include{mydefs}

\newcommand\spart{\;\raise1.0pt\hbox{/}\hskip-6pt\partial}
\newcommand\spartb{\;\overline{\raise1.0pt\hbox{/}\hskip-6pt\partial}}

\newcommand\ees{\end{eqnarray}}
\newcommand\bees{\begin{eqnarray}}

\newcommand{\be}{\begin{equation}}
\newcommand{\ee}{\end{equation}}

\newcommand{\beqa}{\begin{eqnarray}}
\newcommand{\eeqa}{\end{eqnarray}}

\newcommand{\bx}{{\bm{x}}}

\newcommand{\bee}{{\bm{e}}}

\newcommand{\dd}{\text{d}}

\newcommand{\nn}{\nonumber}

\newcommand{\obs}{_{\rm O}}
\newcommand{\Gal}{_{\rm G}}

\newcommand{\ra}{\rightarrow}

\graphicspath{ {./Graphs/} }
\begin{document}
\title{The signal of the stochastic gravitational wave background and the angular correlation of its energy density}
\author{Giulia Cusin}
\email{giulia.cusin@unige.ch}
\affiliation{D\'epartement de Physique Th\'eorique and Center for Astroparticle Physics, Universit\'e de Gen\`eve, 24 quai Ansermet, CH--1211 Gen\`eve 4, Switzerland}
\author{Cyril Pitrou}
\email{ pitrou@iap.fr}
\affiliation{Institut d'Astrophysique de Paris, CNRS UMR 7095, 
           Sorbonne Universit\'e, Institut Lagrange de Paris, 98 bis, Bd Arago, 75014 Paris, France}
\author{Jean-Philippe Uzan}
\email{uzan@iap.fr}
\affiliation{Institut d'Astrophysique de Paris, CNRS UMR 7095, 
           Sorbonne Universit\'e, Institut Lagrange de Paris, 98 bis, Bd Arago, 75014 Paris, France}
\vspace{1 em}
\date{\today}

\begin{abstract}

The gravitational wave radiation emitted by all, resolved and unresolved, astrophysical sources in the observable universe generates a stochastic background. This background has a directional dependence inherited from the inhomogeneities of the matter distribution. This article proposes a new and independent derivation of the angular dependence of its energy density by focusing on the total gravitational wave signal produced by an ensemble of incoherent sources. This approach clarifies the origin of the angular correlation and the relation between the gravitational wave signal that can be measured by interferometers and the energy density of the stochastic background.
\end{abstract}
\maketitle

\section{Introduction}\label{sec_intro}

The superposition of the gravitational wave (GW) radiation emitted by all, resolved and unresolved, astrophysical sources in our universe is at the origin of a stochastic background of gravity waves of astrophysical origin (AGWB). This background has a directional dependence inherited from the inhomogeneities of the matter distribution in the universe, in full analogy with the electromagnetic background of radiation, see e.g. Refs.~\cite{sw67, 1967ApJ...148..377P, Bond:1985pc, Puget:1996fx}.
Moreover, the fact that an emitted GW signal propagates in an inhomogeneous universe, gives an additional effect similar to lensing in optics. 

In a previous analysis~\cite{Cusin:2017fwz}, we provided an expression for the AGWB energy density $\dd^3\rho_{\rm GW}({\bm e}_{\obs},\nu_{\obs})/(\dd^2{\bm e}_{\obs}\dd\nu_{\obs})$ observed in a solid angle $\dd^2{\bm e}_{\obs}$ around a direction  ${\bm e}_{\obs}$,  for an observed frequency $\nu_{\obs}$. This expression is similar to the Sachs-Wolfe formula~\cite{sw67} for the Cosmic Microwave Background  (CMB) temperature anisotropy. It relies on an energetic analysis and on a coarse-graining from astrophysical to galactic and then cosmological scales so that the observed GW flux per units of solid angle depends on the effective luminosity of all the galaxies in that solid angle. The effective luminosity of a galaxy, being  the sum of the contributions of all the GW sources inside it, depends on the mass of the galaxy but also on many astrophysical parameters such as the star formation rates, the stellar evolution, the formation of binary neutron stars or black hole systems. Hence, the final result for the energy density of the background has an astrophysical dependence and a cosmological dependence through the galaxy distribution and the gravitational potential and velocity field distributions. The fact that these cosmological variables  are correlated on cosmological scales induces an angular correlation of the AGWB energy density, which can be characterized by its angular power spectrum. This quantity also correlates  with other cosmological probes, such as lensing and galaxy number counts.


From an experimental perspective, ground-based interferometers,  such as  LIGO and its advanced configuration (aLIGO), Virgo\footnote{\url{https://www.ego-gw.it/public/about/whatIs.aspx}},  and Pulsar timing arrays, such as the radio telescope Parkes Pulsar Timing Array\footnote{\url{http://www.atnf.csiro.au/research/pulsar/ppta/}} (PPTA), the European Pulsar Timing Array \footnote{\url{http://www.leap.eu.org}} (EPTA), the International Pulsar Timing Array\footnote{\url{http://www.ipta4gw.org}} (IPTA) and the North American Nanohertz Observatory for Gravitational Waves\footnote{\url{http://www.nanograv.org}} (NANOGrav), do not directly measure the AGWB energy density but observable quantities related to the strain (e.g. GW signal of a given frequency, GW polarization, phase differences ...). 
 In Ref.~\cite{TheLIGOScientific:2016dpb}  a search for the isotropic stochastic GW background has been performed using data from Advanced LIGO's first observing run. The  total GW density parameter, i.e. the energy density in units of the critical density $\rho_c=8\pi G/(3H_{0}^2)$, ``is constrained to be $\Omega_{\rm GW}<1.7 \times 10^{-7}$ with 95\% confidence, assuming a flat energy density spectrum in the most sensitive part of the LIGO band (20-86 Hz)".  %
At low frequencies, Pulsar Timing Arrays give different bounds, see  Refs.~\cite{Shannon:2013wma, Shannon:2015ect}, which are still under debate \cite{Arzoumanian:2018saf}. 
The possibility of measuring and mapping the gravitational wave background is discussed in Refs.~\cite{Allen:1996gp, Cornish:2001hg, Mitra:2007mc, Thrane:2009fp, Romano:2015uma, Romano:2016dpx} while the description of the different methods  which can be used by LIGO and LISA (Laser Interferometer Space Antenna) to reconstruct an angular resolved map of the sky can be found in Ref.~\cite{TheLIGOScientific:2016xzw}.   An analogous discussion for  Pulsar Timing Arrays  is presented in Refs. \cite{Mingarelli:2013dsa,Taylor:2013esa, Gair:2014rwa,Taylor2015, Anholm:2008wy}. 

An introduction to the different astrophysical sources contributing to this background can be found in Refs.~\cite{Buonanno:2014aza, Allen:1996vm,Regimbau:2011rp} while motivations for  direct searches for a stochastic GW background, both isotropic and anisotropic,  can be found in  Ref.~\cite{Mazumder:2014fja}.

The AGWB signal is usually characterized in terms of the amplitude for a given polarization $A$, $h_A({\bm e}_{\obs},\nu_{\obs})$. A natural question concerns the way to relate this observed GW signal to the AGWB energy density and its correlation function, computed in Ref.~\cite{Cusin:2017fwz}. The goal of this article is to make the relation between the two approaches explicit and to clarify some ambiguities.

This work gives an independent derivation of the angular dependence of the AGWB energy density and extends our former analysis of Ref.~\cite{Cusin:2017fwz}. It explicitly shows why it is correlated even though the AGWB signal is the superposition of incoherent signals. It also paves the way to the reflection on the possibility to measure the stochastic GW background and on the different methods which can be used to achieve this goal.\\

The article is organized as follows. After a summary of some general definitions in Section~\ref{general}, Section \ref{characterization} defines the energy density of the AGWB per units of solid angle as the flux of GW from astrophysical sources that we receive in a given direction and recalls the expression derived in Ref.~\cite{Cusin:2017fwz}. Section~\ref{propagation}  focuses on the propagation of GW in a curved spacetime, in the eikonal approximation. This approach is very similar to the one used in optics. To conclude, in Section~\ref{final} we start from the relation between the observed AGWB energy density and the observed GW signal and we show how the first quantity can be expressed as a function of the signals at emission. Before going to  technical details, Section~\ref{heuristic} proposes an heuristic explanation of the relation between  the AGWB energy density and the observed signal.  It also explains why despite the fact that both the signal and the energy density are stochastic quantities, only the latter has a non-vanishing two-point correlation function. 

To clarify the notation and vocabulary used, we emphasize that we use the expression \emph{GW signal} for the strain or any linear response to the strain and AGWB for the GW energy density, which is quadratic in the strain. 

\section{Heuristic argument}\label{heuristic}

For the sake of the argument, we model galaxies as point-like sources each one emitting GW of amplitude $h_i$. In fact, this signal is given by the incoherent superposition of all of the GW emitted by the sources inside the galaxy. However, neglecting this additional complication does not alter the main argument. Later in this paper we will refine our description. 

The amplitude of the GW signal measured in the direction ${\bm e}_{\obs}$ and in the solid angle $\dd^2{\bm e}_{\obs}$ per frequency $\nu_{\obs}$ is simply the sum of the signals emitted by the sources contained in a bundle of the observer past light cone around the direction of observation. Schematically it is of the form 
\be\label{prima}
h_{\rm obs}({\bm x}_{\obs},t_{\obs} ,{\bm e}_{\obs}; t) \propto \sum_i^{N({ \bm e}_{\obs})} h_i[P_{\rm em}({\bm x}_{\obs}, t_{\obs}, \bee_{\obs}), t]e^{i\varphi_i}\,,
\ee
where $t_{\obs}$ is the cosmic time today (i.e. at the observer position) and $t$ stands for the time measured in the laboratory. $P_{\rm em}({\bm x}_{\obs}, t_{\obs}, \bee_{\obs})$ is the emission point and its coordinates are related by a null geodesic to the observer space-time position $({\bm x}_{\obs}, t_{\obs})$ and to the direction of observation $\bee_{\obs}$. The number of sources along this line of sight is given by $N({\bm e}_{\obs})$, which is a stochastic variable related to the source distribution.\footnote{In a more refined description, this sum can be thought as an integral over the light cone parameterized, e.g. by the redshift, so that the stochastic variable will simply be the number of sources in the beam for a given redshift  bin.} In Eq. (\ref{prima}) the signal emitted by a generic $i$-source has an associated  random phase $\varphi_i$ and the total signal received in a given direction (left hand side of Eq. (\ref{prima})) is made up by the incoherent superposition of all these signals. 

This GW signal can in principle be measured by interferometers, and it is related to the energy density per unit of solid angle by
\begin{align}\label{seconda}
&\frac{\dd^2\rho_{GW}}{\dd^2\bee_{\obs}}({\bm x}_{\obs}, t_{\obs},  \bee_{\obs})\nn\\
& \propto [\dot{h}_{\rm obs}(t_{\obs}, \bx_{\obs}, \bee_{\obs}; t) \dot{h}_{\rm obs} (t_{\obs}, \bx_{\obs}, \bee_{\obs}; t) ]\,,
\end{align}
where a dot denotes a derivative with respect to $t$ and the square brackets refer to a time average on a time scale larger than the typical period of the signal. Let us replace Eq.~(\ref{prima}) into Eq.~(\ref{seconda}) and use the fact that when we sum over a large number of products of incoherent signals (with random phases) non-vanishing contributions to the sum are coming just from products of signals with $\varphi_i=\varphi_j$. Explicitly
\begin{align}\label{terza}
&\frac{\dd^2\rho_{GW}}{\dd^2\bee_{\obs}}({\bm x}_{\obs}, t_{\obs},  \bee_{\obs})\nn\\
& \propto \sum_{i}^{N(\bee_{\obs})} \sum_{j}^{N(\bee_{\obs})}\left[\dot{h}_i[P_{\rm em}, t] \dot{h}^*_j[P_{\rm em}, t]\right]e^{i(\varphi_i-\varphi_j)}\nn\\
&\propto \sum_{i}^{N(\bee_{\obs})}\left[\dot{h}_i[P_{\rm em}, t] \dot{h}^*_i[P_{\rm em}, t]\right]\,,
\end{align}
where we have used the shortcut notation $P_{\rm em}=P_{\rm em}[t_{\obs}, \bx_{\obs}, \bee_{\obs}]$ and going to the last line we have used the fact that only products of coherent signals give non-vanishing contributions to the sum. It follows that the total energy density per units of solid angle is equal to the sum of contributions to the energy density from single sources contained in the solid angle. Explicitly 
\be\label{Icangivestupidnamesaswell1}
\frac{\dd^2\rho_{GW}}{\dd^2\bee_{\obs}}({\bm x}_{\obs}, t_{\obs}, \bee_{\obs}) \propto \sum_i^{N({\bm e}_{\obs})} \frac{\dd^2\rho_{GW, i}}{\dd^2\bee_{\obs}}[P_{\rm em}({\bm x}_{\obs},t_{\obs}, \bee_{\obs})]\,.
\ee
This quantity is a stochastic variable since $N({\bm e}_{\obs})$ is a stochastic variable. 
It follows that the correlation function between different directions
\be\label{Corr}
C(\bee_{\obs}\cdot\bee_{\obs}') = \left\langle \frac{\dd^2\rho_{GW}}{\dd^2\bee_{\obs}}(\bee_{\obs})  \frac{\dd^2\rho_{GW}}{\dd^2\bee_{\obs}'}(\bee_{\obs}') \right \rangle\,,
\ee
inherits the stochastic properties of the variable $N({\bm e}_{\obs})$ as it will be demonstrated in section \ref{final}.  In Eq.~(\ref{Corr}), the angular brackets stand for an average on the cosmological stochastic variables. From the mapping of $\dd^2\rho_{GW}/\dd^2\bee_{\obs}$ obtained in principle thanks to GW radiometry~\cite{Mitra:2007mc}, we can form an estimator of (\ref{Corr}), exactly like the $C_\ell$ of the CMB are estimated from its observed intensity map.

We observe that  the total signal received in a given direction, Eq. (\ref{prima}), is also stochastic since the number of sources $N(\bee_{\obs})$ is a stochastic quantity. However, the two-point correlation function between GW signals from different directions is vanishing due to the fact that when summing products of signals from single sources, only signals with coherent phases (i.e. coming from the same source) give non-vanishing contributions to the sum. Explicitly, one has
\begin{align}
&\langle h_{\rm obs}({\bm x}_{\obs},t_{\obs} ,{\bm e}_{\obs}; t) h_{\rm obs}({\bm x}_{\obs},t_{\obs} ,{\bm e}'_{\obs}; t)\rangle \nn\\
&\propto \delta^2(\bee_{\obs}-\bee_{\obs}') \langle \sum_i^{N(\bee_{\obs})} \Big| h_i[P_{\rm em}(t_{\obs}, \bx_{\obs}, \bee_{\obs}), t]\Big|^2\rangle\,.
\end{align}
It follows that the good quantity to describe the anisotropies of the GW background is not the GW signal received in a given direction (which is a stochastic variable with vanishing correlation function for different directions), but the energy density of the background. This quantity is quadratic in the signal and therefore does not depend on random phases. Even if the GW sources are uncorrelated due to their incoherent nature, the energy density of the GW background they collectively produce is correlated. 

Naively, we could conclude that the correlation function $C$ in Eq. (\ref{Corr}) is related to the correlation function of the number of galaxies weighted by the GW luminosity of the galaxies. This description is indeed simplistic but it explains clearly the origin of the correlation. In order to make it more rigorous, in the rest of this paper we shall
\begin{enumerate}
\item define the averages $[\dots]$ acting on the GW and $\langle\dots\rangle$ acting on the cosmological variables;
\item relate the GW signal emitted by a galaxy to the observed signal. This requires to study the propagation of the GW in a perturbed cosmological spacetime. Both the geodesic equation in the eikonal limit and the Sachs equation for GW will be needed in order to determine the evolution of the amplitude of the waves;
\item determine the GW emitted by a galaxy as a function of the sources it contains (BH, NS binary systems, etc.). This will define the GW luminosity of the galaxy which will depend on the parameters of the galaxy (mass, metallicity,...) and on its evolution (star formation rate, stellar evolution,...) as a function of the signals emitted by all the sources inside the galaxy. 
\end{enumerate}
Hence, the final result for the received signal as a function of the emitted ones indeed depends on the galaxy number density, but also on the gravitational potential and on the velocity field since they enter in the geodesic and Sachs equations.

The fact that the GW signal of the AGWB is not correlated whereas the energy density does correlate is not specific to a GW background: exactly the same situation is realized for its electromagnetic counterpart. For example, for the cosmological background of electromagnetic radiation, the CMB, the {\it electric field} that we receive from different directions plays an analogous role to the GW signal and it is an uncorrelated field. On the other side, the analogous of the energy density of the AGWB is the CMB {\it intensity}, which is proportional to the square of the field and is characterized by a non-vanishing two-point correlation function. The situation for the 21cm line diffuse background is much more similar to the AGWB since its intensity mapping is performed in radioastronomy, that is from the measurement of the electric field through networks of radio-antenna, as e.g. in the LOFAR\footnote{\url{http://www.lofar.org/}} experiment.

A derivation of the AGWB energy density based on an energetic analysis was presented in Ref.~\cite{Cusin:2017fwz}. We propose now an alternative geometrical  derivation of this result, following the approach sketched in this section.

\section{General definitions}\label{general}

This section details the definitions of the averages used in our analysis and then recalls some textbook results on the coarse-grained approach to GW propagation and on the expression for the flux of GW. We mostly follow  Refs.~\cite{Maggiore:1900zz,Isaacson:1967zz,1961RSPSA.264..309S}. 

\subsection{Averages}

We have seen that two different averages appear in our approach. They are different in nature and for the variables on which they act.
\begin{enumerate}
\item The symbol $[\cdots]$ denotes the average entering in the definition of the flux/energy density of GW.  GW vary on a time-scale much smaller than typical astrophysical scales and also much shorter than the characteristic time-scale of the experiment. Given a  physical system, e.g. the GW interferometer, characterized by a given observable quantity $\mathcal{A}$,  we denote by $[\mathcal{A}]$ the time average of $\mathcal{A}$ on a time-interval $T_{\obs}$ much larger that the typical time-scale on which $\mathcal{A}$ varies,
\be\label{At}
[\mathcal{A}(t)]\equiv\frac{1}{T_{\obs}}\int_{0}^{T_{\obs}} \dd t\mathcal{A}(t)\,.
\ee
\item The symbol $\langle \cdots \rangle$ denotes the ensemble average over stochastic initial conditions of the cosmological variables, such as density field, gravitational potential or velocity field. The stochasticity of these variables is inherited from their quantum origin during inflation. This average is the usual ensemble average used in cosmology to compute correlation functions and angular power spectra of cosmological observables~\cite{PeterUzan2005}. If $\mathcal{B}$ is a stochastic quantity, to compare the statistical properties of its observed distribution and the theoretically predicted ones inside a given model, it is necessary to introduce the spatial analog of the ergodic hypothesis.\footnote{The observed distribution is obtained by performing a sky-average of  a single realization, while  the theoretical one is obtained from an ensemble average on some stochastic initial conditions in the frame of a model. This will give rise to an irreducible cosmic variance.}
 \end{enumerate}
We emphasize that if $\mathcal{B}$ is a stochastic quantity, then the time average $[\mathcal{B}]$ is still stochastic.  To avoid confusion we will refer to the average $\langle \dots \rangle$ as \emph{stochastic} ensemble average. 

As we have explained in the previous section, the energy density of the GW background is naturally defined in terms of the average $[\dots]$ of the GW signal, which is still a stochastic field.

\subsection{Coarse-grained form of Einstein equations}

Let us now summarize the standard GR results on the propagation of energy carried by GW. GW are perturbations over some curved, dynamical, background metric $\bar{g}_{\mu\nu}$ so that
\be
g_{\mu\nu}=\bar{g}_{\mu\nu}+ h_{\mu\nu}\,,\qquad |h_{\mu\nu}|\ll 1\,.
\ee
A natural splitting between the space-time background and gravitational waves arises when there is a clear separation of scales.  In particular, a natural distinction can be made in frequency space, if $\bar{g}_{\mu\nu}$ has frequencies up to a maximum value $\nu_B$ while $h_{\mu\nu}$ is picked around frequency $\nu$ such that 
\be
\nu\gg \nu_B\,.
\ee
In this case $h_{\mu\nu}$ is a high-frequency perturbation of a static or slowly varying background.

The Einstein equations can then be expanded up to quadratic order in $h_{\mu\nu}$ and split in high- and low-frequency modes to define, after averaging them over time, i.e. using the average $[\cdots]$,  the GW energy tensor $t_{\mu\nu}$, see Ref.~\cite{Maggiore:1900zz}.   
This \emph{coarse-grained} form of the Einstein equations~\cite{Maggiore:1900zz} determines the dynamics of $\bar{g}_{\mu\nu}$ in terms of the low-frequency part of the energy momentum tensor of matter $\bar{T}_{\mu\nu}$ and of a tensor $t_{\mu\nu}$ which does not depend on the external matter but only on the gravitational field itself at quadratic order in $h_{\mu\nu}$. It can be checked that the following conservation equation holds 
\be\label{Bianchi}
\bar{D}^{\mu}(\bar{T}_{\mu\nu}+t_{\mu\nu})=0\,,
\ee
where $\bar{D}^{\mu}$ denotes the covariant derivative with respect to the background metric. 
The high-frequency part of the Einstein equations then implies  that (see e.g. Ref.~\cite{Maggiore:1900zz} for details)
 \be\label{prop}
 \bar{D}^{\rho} \bar{D}_{\rho} \bar{h}_{\mu\nu}=0\,,
 \ee
at leading order in $\nu_B/\nu$ once we adopt the Lorentz gauge, defined by the condition
 \be\label{gauge}
 \bar{D}^{\nu} \bar{h}_{\mu\nu}=0\,,
 \ee
where we have defined 
 \be
 \bar{h}_{\mu\nu} \equiv h_{\mu\nu}-\frac{1}{2}\bar{g}_{\mu\nu} h\,,
 \ee
 with $h=h_{\mu\nu}\bar{g}^{\mu\nu}$. Equation (\ref{prop}) together with the gauge condition (\ref{gauge}), determines the propagation of GW on a curved background in the limit $\nu_B/\nu\ll1$. When specialized to a perturbed Friedmann-Lema\^{\i}tre spacetime, this equation allows one to characterize the effect of the large scale structures on the wave form, see  e.g. Ref.~\cite{Bonvin:2016qxr}. \\ 
 

 
 \subsection{The energy-momentum tensor of GW}\label{TTs}

Let us now present the explicit form of $t_{\mu\nu}$  in order to make its physical interpretation clear. 

Far from the sources (e.g. at the position of the detector) the background spacetime is well-approximated by a Minkowski spacetime, i.e. $\bar{g}_{\mu\nu}=\eta_{\mu\nu}$ and  $\bar{D}_{\mu}\rightarrow \partial_{\mu}$ (in Minkowskian coordinates). It follows that the equation describing the GW propagation, Eq.~(\ref{prop}),  becomes
\be\label{plane}
\Box h_{\mu\nu}=0\,,
\ee
where $\Box$ is the flat-space d'Alembertian, and the Lorentz-gauge condition (\ref{gauge}) reduces to
\be
\partial^{\mu}h_{\mu\nu}=0\,.
\ee
This gauge condition is not spoiled by a further coordinate transformation $x^{\mu}\rightarrow x^{\mu}+\xi^{\mu}$ with $\Box \xi_{\mu}=0$. This residual gauge freedom still allows one to impose $\bar{h}=0$ and $h^{0i}=0$,  so that the Lorentz condition implies in particular $\partial^0 h_{00}=0$. This leads to the conditions
\be\label{TT}
h^{0\mu}=0\,,\quad h^i_i=0\,, \quad \partial^j h_{ij}=0\,,
\ee
which completely fix the transverse-traceless (TT) gauge. The tensor $t_{\mu\nu}$ far from the sources takes the form 
\be\label{tmunu}
t_{\mu\nu}=\frac{c^4}{32\pi G} [\partial_{\mu}h_{\alpha\beta} \partial_{\nu} h^{\alpha\beta}]\,.
\ee
It can be checked that this object is invariant under a linearized gauge transformation $h_{\mu\nu}(x)\ra h_{\mu\nu}(x)-(\partial_{\mu} \xi_{\nu}+\partial_{\nu}\xi_{\mu})$. As a consequence, $t_{\mu\nu}$ depends only on the physical modes $h_{ij}^{TT}$ and one can just replace the metric $h_{\mu\nu}$ in Eq.~(\ref{tmunu}) with the metric in TT gauge. In particular, the gauge invariant energy density is given by 
\be\label{t00}
t^{00}=\frac{c^2}{32\pi G} [\dot{h}^{TT}_{ij} \dot{h}^{TT}_{ij}]\,,
\ee
where the dot denotes $\partial_t=(1/c) \partial_0$. To conclude,  far from the sources where $T_{\mu\nu}\ra 0$,  Eq.~(\ref{Bianchi}) reduces to 
\be\label{tm}
\partial^{\mu}t_{\mu\nu}=0\,.
\ee

\subsection{The energy flux}\label{mmmm}

The energy flux is the energy of GW flowing per unit of time through a unit surface at a large distance from the source. From the conservation equation~(\ref{tm}) for the energy momentum tensor, it follows that 
\be\label{R0}
\int_V \dd^3\bx (\partial_0 t^{00}+\partial_i t^{i0})=0\,,
\ee
where $V$ is a spatial volume in the far region, bounded by a surface $S$. The GW energy inside the volume $V$ is
\be
E_V=\int_V \dd^3\bx\, t^{00}\,,
\ee
so that Eq.~(\ref{R0}) becomes
\be
\frac{1}{c}\frac{\dd E_V}{\dd t} =-\int_V \dd^3\bx\,\partial_i t^{0i}=-\int_S \dd A\, e_i t^{i0}\,,
\ee
where $e^i$ is the outer normal to the surface and $\dd A$ is the surface element.\footnote{More precisely, we take as volume $V$ a spherical shell centered on the source but far away from it, in such a way that both its inner source and its outer source, $S_1$ and $S_2$ respectively are in the wave region. The time derivative of $E_V$ is given by the sum of two contributions: the energy flowing in through  $S_1$ minus the energy flowing out from $S_2$. We are interested in the energy flux at a given distance from the source (e.g. in the energy flowing through a unit surface of our detector) which for definiteness we choose to be on the outer surface $S_2$ so in the following we take $S=S_2$.} Let $S$ be a spherical surface at large distance $r$ from the source, then $\dd A=r^2\dd\Omega$ and ${\bm e}=\hat{\bm r}$ is the unit vector in the radial direction.  Thus, we get
\be
\frac{\dd E_V}{\dd t}=-c \int \dd A t^{0r}\,,
\ee
where 
\be
t^{0r}=\frac{c^4}{32\pi G} [\partial^0 h_{ij}^{TT} \frac{\partial}{\partial r} h_{ij}^{TT}]\,.
\ee
Using that far from the source $\partial_r h_{ij}^{TT}(t, r)=-\partial_0 h_{ij}^{TT}(t, r)+\mathcal{O}(1/r^2)$ $=$
$\partial^0h_{ij}^{TT}(t, r)+\mathcal{O}(1/r^2)$,
we get 
\be
\frac{\dd E_V}{\dd t}=- c\int \dd A t^{00}\,.
\ee
The  fact that $E_V$ decreases means that the outward-propagating GW carries away an energy flux 
\begin{align}\label{flux1}
&\frac{\dd^2E}{\dd A \dd t}(t_{\obs}, \bx_{\obs},  \bee_{\obs})=c t^{00}\\
&=\frac{c^3}{32\pi G} [\dot{h}_{ij}^{TT}(t_{\obs}, \bx_{\obs}, \bee_{\obs}; t) \dot{h}_{ij}^{TT} (t_{\obs}, \bx_{\obs}, \bee_{\obs}; t) ]\nn\,,
\end{align}
where we have explicitly indicated the dependence on the observer space-time position $(t_{\obs}, \bx_{\obs})$ (where the unit surface $\dd A$ is located) and on the direction of observation $\bee_{\obs}$.  This energy flux has dimension of $[\text{mass}^4]$. We observe that, being defined as an average over the time of observation $t$, the left hand side of Eq. (\ref{flux1}) does not depend on $t$. 

It is useful to introduce a polarization basis $\{\epsilon_{ij}^{+}, \epsilon_{ij}^{\times}\}$ satisfying $\epsilon^A_{ij}(\bee)\epsilon_B^{ij}(\bee)=2\delta^A_B$ so that the two degrees of freedom of the GW are decomposed as
\be\label{farf}
h_{ij}^{TT} (t_{\obs}, \bx_{\obs}, \bee_{\obs}; t)=\sum_{A=(+, \times)}h_{A}\epsilon^A_{ij}\,,
\ee
in terms of which Eq.~(\ref{flux1}) becomes 
\begin{align}\label{flux33}
& \frac{\dd^2E}{\dd A \dd t}(t_{\obs}, \bx_{\obs}, \bee_{\obs})=c \rho_{GW}(t_{\obs}, \bx_{\obs}, \bee_{\obs})\\
&=\frac{c^3}{16\pi G} \sum_{A=(+, \times)}[\dot{h}_A(t_{\obs}, \bx_{\obs}, \bee_{\obs}; t) \dot{h}_{A} (t_{\obs}, \bx_{\obs}, \bee_{\obs}; t) ]\nn\,,
\end{align}
where we have denoted as $ \rho_{GW}(t_{\obs}, \bx_{\obs}, \bee_{\obs})$ the energy density of the source we are considering, received in the direction $\bee_{\obs}$. 

If we consider the contribution of several sources located in an infinitesimal solid angle $\dd^2 \bee_{\obs}$, they give a total observed  amplitude (of a given polarization) $\dd^2{h}^{\text{tot}}_A$. The corresponding infinitesimal energy density 
in the solid angle $d^2\bee_{\obs}$ is given by 
\begin{align}\label{flux33infinitesimal}
&\dd^2 \rho_{GW}(t_{\obs}, \bx_{\obs}, \bee_{\obs}) =\frac{c^2}{16\pi G} \times\\
&\quad\sum_{A=(+, \times)}[\dd^2 \dot h^{\text{tot}}_A(t_{\obs}, \bx_{\obs}, \bee_{\obs}; t)\dd^2 \dot h^{\text{tot}}_{A} (t_{\obs}, \bx_{\obs}, \bee_{\obs}; t) ]\nn\,.
\end{align}

\section{Characterization of a GW background}\label{characterization}

\subsection{Definitions}\label{defi}

The background of GW of astrophysical origin can be characterized in terms of its energy density defined as 
\be
\rho_{GW}(t_{\obs}, \bx_{\obs})=\int \dd^2\bee_{\obs}\, \frac{\dd^2\rho_{GW}}{\dd^2\bee_{\obs}}(t_{\obs}, \bx_{\obs}, \bee_{\obs})\,,
\ee
where the integrated quantity on the right hand side is the energy density of the background per unity of solid angle, which is related to the total amplitude received through Eq.~(\ref{flux33infinitesimal}). 
%

\subsection{Our parametrization}

In Ref.~\cite{Cusin:2017fwz}, we derived an analytic expression for the energy density of GW in terms of the sum of the fluxes from galaxies located in the solid angle around the direction of observation, integrated along the line of sight  (for an alternative tentative based on the Boltzmann equation see Ref.~\cite{Contaldi:2016koz}). The flow received from a galaxy was expressed as a function of the effective luminosity of the galaxy. The effective luminosity of a galaxy was then written by considering the contributions from the different GW sources it contains. In other words, the idea underlying our approach was to introduce different scales in the problem and to coarse-grain from one to the other. This procedure allowed us to write the energy density of the GW background in terms of quantities defined on local scales of single GW sources inside a galaxy. 

In order to obtain a parametrization for the GW signal that we receive from all resolved and unresolved GW sources, we work in the same framework proposed in Ref.~\cite{Cusin:2017fwz}. We distinguish three scales: 
\begin{itemize}
\item \emph{cosmological scale}. The observer measures a GW signal in a  solid angle $\dd^2\bee_{\obs}$ around a direction $\bee_{\obs}$. The angular resolution of the observer is such that we assume galaxies to be point-like sources emitting GW and comoving with the cosmic flow.
\item \emph{galactic scale}. A galaxy is described by a set of parameters  $\theta_{\Gal}$ such as its mass, mean metallicity, etc. We associate to each galaxy an effective GW total signal given by the superposition of the GW signals emitted by all the single GW sources it contains. 
\item \emph{astrophysical scale}. This is the local scale of single GW sources. 
\end{itemize}
It follows that the observed GW signal depends on sub-galactic parameters (properties of the evolution of binary systems, production of GW by astrophysical sources,...), galactic parameters (star formation rate, total mass, evolution of the metallicity,...) and cosmological parameters (distribution of the gravitational potential, number density of galaxies, velocity fields).

\subsection{Summary of our previous result}

In Ref.~\cite{Cusin:2017fwz}, we found that
 \begin{widetext} \be\label{pippo2}
 \frac{\dd^3\rho_{GW}}{\dd\nu_{\obs}\dd^2\bee_{\obs}}(t_{\obs}, \bx_{\obs}, \bee_{\obs}, \nu_{\obs})=\frac{1}{4\pi}\int \dd\lambda \int \dd\theta_{\Gal} \frac{\sqrt{p_{\mu}(\lambda) p^{\mu}(\lambda)}}{\left[1+z_{\Gal}(\lambda)\right]^3}\,n_{\Gal}\left[x^{\mu}(\lambda), \theta_{\Gal}\right] \mathcal{L}_{\Gal}(\nu_{\Gal}, \theta_{\Gal})\,,
 \ee
   \end{widetext}
 where $\mathcal{L}_{\Gal}(\nu_{\Gal}, \theta_{\Gal})$ is the effective luminosity of a galaxy and $\nu_{\Gal}$ is the effective frequency of the galaxy.  In this expression, we integrate along the line of sight, parameterized by the affine parameter $\lambda$. Each galaxy in the solid angle of observation is characterized by a set of parameters $\theta_{\Gal}$ (e.g. mass, metallicity,...). The quantities $z_{\Gal}$ and $n_{\Gal}$ correspond to the redshift and number density of galaxy, respectively, while $p^\mu$ is the spatial projection of the wave-vector [see Sec.~\ref{sight} below for detailed definitions].
%
 

\section{GW propagation in a universe with structures}\label{propagation}

This section describes the propagation of a GW signal in a generic curved spacetime in the eikonal approximation. Our final goal is to express the GW signal that we receive from a given direction as a function of the one at emission. Plugging this result in Eq. (\ref{flux33infinitesimal}) we will thus find an expression for the energy density of the AGWB as a function of emitted GW signals. 

\subsection{Eikonal approximation to GW propagation}\label{eikonal}

Our approach follows the standard eikonal approximation in geometric optics~\cite{DeruelleBook}. This approximation holds for wavelengths $\lambdabar$ much smaller than the other typical length-scales in the problem, i.e. $\lambdabar \ll L_B$ where $L_B$ is the typical length-scale of variation of the background geometry and $\lambdabar \ll L_c$, where $L_c$  is the characteristic length-scale over which the amplitude, polarization and wavelength of the field change substantially. In particular, $\lambdabar$ has to be smaller than the curvature radius of the wavefront.   The eikonal approximation consists in looking for solutions of the wave equation with a phase $\theta$ rapidly varying, i.e. $\theta$ varies on a scale $\lambdabar$, while the amplitude and polarization of the wave change on a scale $L_c$, so it is slowly varying. To perform the expansion systematically, it is convenient to expand the GW as
\be
h_{\mu\nu}(x)=\left[H_{\mu\nu}(x)+\varepsilon B_{\mu\nu}(x) +\dots\right]e^{i\theta(x)/\varepsilon}\,,
\ee
where $\varepsilon$ is a fictitious parameter to be finally set equal to unity.\footnote{If a term has a factor $\varepsilon^n$ attached, it is of the order $(\lambdabar/L)^n$, where $L$ is the smallest scale between $L_B$ and $L_c$.} We emphasize that an expansion of this form is just an ansatz and its validity is verified by substituting it in the equations. 

Defining 
\be
k_{\mu}=\bar{D}_{\mu}\theta=\partial_{\mu}\theta\,,
\ee
and $H_{\mu\nu}=H \epsilon_{\mu\nu}$ with the polarization tensor satisfying $\epsilon^*_{\mu\nu}\epsilon^{\mu\nu}=1$, at leading order in $\varepsilon$, Eqs.~(\ref{prop}) and (\ref{gauge}) describing the propagation of GW on a curved background give, respectively 
\bees
\label{geodesic}\bar{g}_{\mu\nu}k^{\mu} k^{\nu}&=&0\,,\label{111}\\
\epsilon^{\mu\nu}k_{\mu}&=&0\,\label{66}.
\ees
From Eq.~(\ref{111}) it follows that $0=\bar{D}_{\nu}(k_{\mu} k^{\mu})=2 k^{\mu} \bar{D}_{\nu} \bar{D}_{\mu}\theta$, i.e.
\be
0= 2 k^{\mu} \bar{D}_{\mu} k_{\nu}\,,\label{112}
\ee  
which is simply the geodesic equation in the space-time with the background metric $\bar{g}_{\mu\nu}$. Equation~(\ref{111}) implies that the curves orthogonal to the surfaces of constant phase (the rays in the geometric optic approximation) travel along the null geodesics of $\bar{g}_{\mu\nu}$. 

To next-to-leading order in $\varepsilon$, Eq.~(\ref{prop}) gives 
\bees
k^{\mu} \partial_{\mu} H&=&-\frac{H}{2} \bar{D}_{\mu} k^{\mu}\,, \label{55}\\
k_{\rho} (\bar{D}^{\rho} \varepsilon_{\mu\nu})&=&0\,,\label{parallel}
\ees
while Eq.~(\ref{gauge}) gives an equation for $B_{\mu\nu}$, i.e. a correction to the amplitude and polarization $H_{\mu\nu}=H\epsilon_{\mu\nu}$. Equation~(\ref{66}) shows that the polarization tensor is transverse to the wave vector while Eq.~(\ref{parallel}) expresses the fact that it is parallel transported along a geodesic. Equation~(\ref{55}) can be rewritten as 
\be
\bar{D}_{\mu}(H^2 k^{\mu})=0\,,
\ee
which shows that the current $j^{\mu}=H^2 k^{\mu}$ is conserved. Its conserved charge is the integral of $H^2 k^{0}$ over a constant time hypersurface. Taking into account that each graviton carries an energy $k^0$, it can be verified that $H^2 k^0$  is proportional to the number density of gravitons so that the conserved charge is the number of gravitons.\\

The set of equations~(\ref{111}-\ref{parallel}) for $k^{\mu}\equiv \dd x^{\mu}(\lambda)/\dd\lambda$, $\epsilon_{\mu\nu}=\epsilon_{\mu\nu}[x^{\mu}(\lambda)]$ and $H=H[x^{\mu}(\lambda)]$ can be solved with initial conditions at the observer position $\lambda=\lambda_{\obs}$ and determine the wave signal at the spacetime point  $x^{\mu}(\lambda)$, i.e. $h_{\mu\nu}[x^{\mu}(\lambda)]$.

\subsection{Line of sight approach}\label{sight}

Let us start from the geodesic equation~(\ref{111})-(\ref{112}) describing the evolution of the phase of the GW signal in the eikonal approximation. 

We consider an observer with 4-velocity $u^{\mu}$ ($u_{\mu}u^{\mu}=-1$). At any time, his worldline is the origin of the observer past lightcone containing all observed GW rays. The 4-velocity $u^\mu$ defines a preferred spatial section and the spatial direction of GW propagation, defined as the opposite of the direction of propagation of the signal converging to the observer. It is spanned by the spatial unit vector $e^{\mu}$\,,
\be\label{e.def-e}
e^{\mu}u_{\mu}=0\,,\hspace{1 em}e^{\mu}e_{\mu}=1\,,
\ee
which provides the 3+1 decomposition of the wave 4-vector
\be\label{directionobs}
k^{\mu}=E \left(u^{\mu}-e^{\mu}\right)\,,
\ee
where $E=2 \pi \nu\equiv -u_{\mu}k^{\mu}$ is the cyclic frequency of the GW in the observer's rest frame. The spatial projection of the wave 4-vector is
\be\label{p}
p^{\mu}\equiv \left(g^{\mu\nu}+u^{\mu}u^{\nu}\right) k_{\nu}=-E e^{\mu}\,. 
\ee
The redshift $z_{\Gal}$ of a source $G$  is defined from the ratio between the emitted frequency $\nu_{\Gal}$ in the source's rest frame and the observed frequency in the observer's rest frame $\nu_{\obs}$, i.e.
\be\label{redshift}
1+z_{\Gal}\equiv \frac{\nu_{\Gal}}{\nu_{\obs}}=\frac{u_{\Gal}^{\mu}\,k_{\mu}(\lambda_{\Gal})}{u_{\obs}^{\mu}\,k_{\mu}(\lambda_{\obs})}\,,
\ee
where $u^{\mu}_{\Gal}$ is the 4-velocity of the source and $u^{\mu}_{\obs}$ is the 4-velocity of the observer. The source $G$ located at a redshift $z_{\Gal}$ is emitting GW with a given frequency spectrum. From the definition~(\ref{redshift}), it follows that the frequency measured in $O$, $\nu_{\obs}$, is related to the frequency at the emission, $\nu_{\Gal}$, by 
\be\label{nus}
\nu_{\Gal}=(1+z_{\Gal})\nu_{\obs}\,.
\ee
Since $x^{\mu}(\lambda)$ is the worldline of a graviton which intersects the worldline of the observer at the time of observation, it follows that
\be\label{11}
x^{\mu}(\lambda_{\obs})=x_{\obs}^{\mu}\,,\hspace{1 em} \left. \frac{\dd x^{\mu}(\lambda)}{\dd\lambda}\right|_{\lambda=\lambda_{\obs}}=E_{\obs}(u_{\obs}^{\mu}-e_{\obs}^{\mu})\,.
\ee
Therefore, $x^{\mu}(\lambda)$ is a function of the direction of observation and of 4-position of the observer, i.e. $x^{\mu}(\lambda)=x^{\mu}(\lambda, e^{\mu}_{\obs}, x^{\mu}_{\obs})$. In the following, to make the notation compact, the dependence on $e^{\mu}_{\obs}$ and $x^{\mu}_{\obs}$ will be understood. 

\subsection{Evolution of the GW amplitude}

To derive the evolution of the GW amplitude from Eq.~(\ref{55}), we study the deformation of a bundle of null geodesics propagating in an inhomogeneous spacetime. As we will show, the physical area of the beam is related to the amplitude of the GW signal in the eikonal approximation. 

Consider a geodesic bundle converging at the observer position in $O$. In $O$, we choose an orthonormal basis $\{k^{\mu}, u^{\mu}, s_1^{\mu}, s_2^{\mu}\}$ where 
\be
k^{\mu}\equiv \frac{\dd x^{\mu}_R}{\dd \lambda}\,,
\ee
is the tangent vector to the null reference-geodesic $x^{\mu}_R$, $u^{\mu}$ is the tangent vector to the observer's worldline and the two spacelike vectors $s_1^{\mu}$ and $s_2^{\mu}$ are spanning the plane perpendicular to the line of sight, i.e.
\begin{align}
&\quad u_{\mu} u^{\mu}=-1\,,\qquad k^{\mu}k_{\mu}=0\,,\nn\\
& s_{\mu}^a s_{b}^{\mu}=\delta^a_b\,,\quad s_a^{\mu}k_{\mu}=s^{\mu}_a u_{\mu}=0\,.
\end{align}

In full analogy with the electromagnetic case~\cite{1992grle.bookS,DeruelleBook}, the Jacobi matrix $\bm{\mathcal{D}}$ describes the propagation of light (GW) beams. The associated deformation matrix is naturally defined by
\be\label{1}
\bm{\mathcal{S}}\equiv \frac{\dd\bm{\mathcal{D}}}{\dd\lambda}\bm{\mathcal{D}}^{-1}\,.
\ee
 It can be shown (see \S~2.3 of Ref.~\cite{Fleury:2015hgz}) that this matrix is symmetric. It is usually decomposed into a trace ,
  \be
 \text{tr}\bm{\mathcal{S}}\equiv2\theta\,,
 \ee
 and a trace-free part introducing the so-called optical scalars. Alternatively,  $\bm{\mathcal{S}}$ can be defined by
\be\label{2}
\bm{\mathcal{S}}_{ab}=s_{a}^{\mu}s_{b}^{\nu} \bar{D}_{\mu} k_{\nu}\,,
\ee
which can be checked to be equivalent to Eq.~(\ref{1}); see Refs.~\cite{Fleury:2015hgz,Fleury:2013sna,Fleury:2014rea}. By decomposing the tensor $\bar{D}_{\mu}k_{\nu}$ over the orthonormal basis $(u^{\mu}, d^{\mu}, s_1^{\mu}, s_2^{\mu})$ and taking the trace, it can be verified that
\be\label{sachs}
\bar{D}_{\mu} k^{\mu}=\text{tr} \bm{\mathcal{D}}\,.
\ee

The physical cross-sectional area of a light beam is defined by
\be
A\equiv \int_{\text{beam}} \dd\xi^1\dd\xi^2=\int_{\text{beam}} \det\bm{\mathcal{D}}\, \frac{\dd\xi_{\obs}^1}{\dd\lambda}\frac{\dd\xi_{\obs}^2}{\dd\lambda}\,.
\ee
For an infinitesimal beam, $\bm{\mathcal{D}}$ can be considered constant in the above integral and the evolution rate of $A$ with the affine parameter reads 
\be
\frac{1}{A}\frac{\dd A}{\dd\lambda}=\frac{1}{\det\bm{\mathcal{D}}}\frac{\dd(\det\bm{\mathcal{D}})}{\dd\lambda}=\text{tr}\left(\frac{\dd\bm{\mathcal{D}}}{\dd\lambda}\bm{\mathcal{D}}^{-1}\right)=\text{tr}\bm{\mathcal{S}}\,.
\ee
Therefore, using Eq.~(\ref{sachs}), it follows
\be\label{555}
\bar{D}_{\mu}k^{\mu}=2\theta=\frac{1}{A} \frac{\dd A}{\dd\lambda}\,.
\ee
Plugging Eq.~(\ref{555}) in Eq.~(\ref{55}) describing the evolution of a  GW amplitude in the eikonal approximation, and after some trivial manipulations, we find 
\be
\frac{\dd}{\dd\lambda}\left(H^2A\right)=0\,.
\ee
Therefore,
\be
H(\lambda_{\obs})=H(\lambda_{\Gal})\sqrt{\frac{A(\lambda_{\obs})}{A(\lambda_{\Gal})}}\,,
\ee
where $A_{\Gal}\equiv A(\lambda_{\Gal})$ is the physical size of the source and $A_{\obs}\equiv A(\lambda_{\obs})$ is the size of the beam measured at the observer position. Using
the distance duality relation (e.g. \S~3.2.4 of Ref.~\cite{Fleury:2015hgz})
\be\label{mist}
D_L=(1+z_{\Gal})\sqrt{\frac{A_{\obs}}{\Omega_{\Gal}}}\,,
\ee
where  $\Omega_{\Gal}$ is the solid angle subtending the surface of the beam at the observer position seen from the source, the amplitude of the GW measured by the observer $O$, $H^{[\Gal,\obs]}$,  can be expressed as a function of quantities at the emission point $G$ as
\bees\label{oem0}
H^{[\Gal,\obs]}&=&H_{\Gal}(\lambda_{\Gal})\sqrt{\frac{A_{\Gal}}{\Omega_{\Gal}}}\frac{(1+z_{\Gal})}{D_L(\lambda_{\Gal})}\,, 
\ees
where $H_{\Gal}$ is the amplitude at emission. 
From Eq.~(\ref{66})  it follows that the polarization of the wave is parallel transported, i.e. there is no polarization mixing during the GW propagation. Moreover, the phase of the wave remains constant along null geodesic, i.e. $\theta(\lambda_{\Gal})=\theta(\lambda_{\obs})$. Going to TT gauge and using the polarization basis introduces in section \ref{defi}, Eq.~(\ref{farf}),  we can write the GW signal of a given polarization received by the observer $O$, $h_{A}^{[\Gal,\obs]}$ in terms of the emitted one $h_{A}^{[\Gal]}$ as 
\bees\label{oem}
h_{A}^{[\Gal,\obs]}&=&|h_{A}^{[\Gal]}|D_L^{\text{prox}}\frac{(1+z_{\Gal})}{D_L(\lambda_{\Gal})}\exp\{i\varphi^{[\Gal]}\}\,,
\ees
where we have used Eq.~(\ref{oem0}) and we have defined $D_L^{\text{prox}}\equiv \sqrt{A_{\Gal}/\Omega_{\Gal}}$ which corresponds to the limit of Eq.~(\ref{mist}) for $z_{\Gal}\ra 0$, i.e. to the luminosity distance measured by an observer in the vicinity of the source. In Eq. (\ref{oem}) $\varphi^{[\Gal]}$ is a random phase that takes into account that all the sources are incoherent. 
\vspace{3 em}

\section{Recovering the energy density}\label{final}

\subsection{Total GW amplitude}

The total GW signal of a given polarization $A=(+, \times)$ received in $\bx_{\obs}$ at time $t_{\obs}$ in the direction $\bee_{\obs}$, coming from the sources located in an observed solid angle $\dd^2{\bf{e}}_{\obs}$ is
\begin{align}\label{master}
&\dd^2 h_{A}^{\text{tot}}(t_{\obs}, \bx_{\obs}, \bee_{\obs}; t)\\
&=\int \dd\lambda \int \dd\theta_{\Gal} \frac{\dd^3\mathcal{N}_{\Gal}[x^{\mu}(\lambda), \theta_{\Gal}]}{\dd\lambda}h_{A}^{[\Gal, \obs]}[x^{\mu}(\lambda), \theta_{\Gal}; t]\,,\nn
\end{align}
where $h_{A}^{[\Gal, \obs]}[x^{\mu}(\lambda), \theta_{\Gal}; t]$ is the GW signal of polarization $A$ that the observer receives from a galaxy $G$ located in $x^{\mu}(\lambda)$. We have explicitly indicated that it depends on the parameters characterizing the galaxy, $\theta_{\Gal}$. The quantity  $\dd^3 \mathcal{N}_{\Gal}[x^{\mu}(\lambda), \theta_{\Gal}]$ represents the number of galaxies with parameters $\theta_{\Gal}$ contained in the physical volume $\dd^3 V$, seen by the observer $O$ under the solid angle $\dd^2{\bf{e}}_{\obs}$. 

We observe that to make contact with the heuristic argument present in section \ref{heuristic}, we can rewrite Eq. (\ref{master}) by substituting Eq. (\ref{oem}) and factorizing out the random phase of the emitted signal as
\begin{align}
&\dd^2 h_{A}^{\text{tot}}(t_{\obs}, \bx_{\obs}, \bee_{\obs}; t)\\
&=\int \dd\lambda \int \dd\theta_{\Gal} \frac{\dd^3\mathcal{N}_{\Gal}}{\dd\lambda}|h_{A}^{[\Gal, \obs]}|[x^{\mu}(\lambda), \theta_{\Gal}; t]e^{i\varphi^{[\Gal]}}\,.\nn
\end{align}
In the situation sketched in the heuristic argument, sources were labelled by a discrete index $i$. The mapping between the toy model of section \ref{heuristic} and the scenario that we are now describing can be made with the substitutions
\begin{align}
i &\rightarrow (\lambda, \theta_{\Gal})\\
P_{\rm em}[t_{\obs}, \bx_{\obs}, \bee_{\obs}]&\rightarrow x^{\mu}(\lambda)\\
\sum_{i}^{N(\bee_{\obs})}&\rightarrow \int d\lambda \int d\theta_{\Gal} \frac{\dd^3\mathcal{N}_{\Gal}}{\dd\lambda}[x^{\mu}(\lambda), \theta_{\Gal}]\\
\varphi_i&\rightarrow \varphi^{[\Gal]}\,.
\end{align}
Since the phases $\varphi^{[\Gal]}$ are random,  the correlator of the GW signal, being a sum of a large number of contributions with random relative phases, is vanishing unless these relative phases vanish. This can happen only when $\bee'_{\obs}=\bee_{\obs}$. In other words, the two-point correlator of the GW signals from different directions is vanishing, explicitly 
\begin{align}
&\langle \dd^2 h_{A}^{\text{tot}}(t_{\obs}, \bx_{\obs}, \bee_{\obs}; t)\dd^2 h_{A}^{\text{tot}}(t_{\obs}, \bx_{\obs}, \bee'_{\obs}; t)\rangle \nn\\
&\qquad \propto \delta^2(\bee_{\obs}-\bee_{\obs}')\,.
\end{align}

In Eq. (\ref{master}), the received signal  from a galaxy can be expressed in terms of the emitted one by using Eq. (\ref{oem}). Moreover, the number of galaxies in an infinitesimal physical volume, $\dd^3\mathcal{N}_{\Gal}$, can be expressed as a product of physical volume and physical galaxy density
\be\label{NG}
\dd^3\mathcal{N}_{\Gal}\left[x^{\mu}(\lambda), \theta_{\Gal}\right]\equiv n_{\Gal}\left[x^{\mu}(\lambda), \theta_{\Gal}\right] \dd^3 V\left[x^{\mu}(\lambda)\right]\,.
\ee
The physical volume $\dd^3V$ is defined as 
\be\label{covV}
\dd^3V=\sqrt{-g} \epsilon_{\mu\nu\alpha\beta}u^{\mu} \dd x^{\nu} \dd x^{\alpha} \dd x^{\beta}\,. 
\ee
To simplify our final result, it is useful to rewrite Eq.~(\ref{covV}) expressing the physical volume element as 
\be\label{Volume}
\dd^3 V\left[x^{\mu}(\lambda)\right]=\dd^2{\bf{e}}_{\obs} D_{\rm A}^2(\lambda) \sqrt{p_{\mu}(\lambda) p^{\mu}(\lambda)} \dd\lambda \,,
\ee
that uses that $\dd^3 V$ is the volume with cross-section $D_{\rm A}^2$ and depth $\sqrt{p_{\mu}(\lambda) p^{\mu}(\lambda)} \dd\lambda =\textcolor{black}{-(u_\mu k^\mu)}\dd\lambda$ along the line of sight, $p^{\mu}$ being defined in Eq.~(\ref{p}). The angular diameter distance is related to the luminosity distance by the reciprocity relation
\be\label{reciprocity}
D_L=(1+z_{\Gal})^2 D_A\,.
\ee

\subsection{Energy density}

The energy density of the background in a given direction is related to the observed GW signal from that direction through Eq. (\ref{flux33infinitesimal}). We plug Eq. (\ref{master}) in Eq. (\ref{flux33infinitesimal}) and we consider that products of signals coming from incoherent sources give a vanishing contribution to the integral. We find
\begin{widetext}
\be\label{Ettore}
\frac{d^2\rho_{GW}}{d^2\bee_{\obs}}(t_{\obs}, \bx_{\obs}, \bee_{\obs})=\frac{c^2}{16\pi G} \int d\lambda\int d\theta_{\Gal} \frac{d^3\mathcal{N}_{\Gal}}{d\lambda d^2\bee_{\obs}}[x^{\mu}(\lambda), \theta_{\Gal}]\sum_A\left[\dot{h}_A^{[\Gal, \obs]}(x^{\mu}(\lambda), \theta_{\Gal}; t)\dot{h}_A^{[\Gal, \obs]}(x^{\mu}(\lambda), \theta_{\Gal}; t)\right]\,,
\ee
\end{widetext}
where the square parenthesis denotes the time average (\ref{At}) over a period of time $T_{\obs}$ much larger than the characteristic period of the GW signal, see section \ref{general} for definitions. To simplify this time average we introduce the Fourier transform with respect to the time at the detector $t$ as
\be\label{Proserpina}
h_A^{[\Gal, \obs]}(x^{\mu}(\lambda), \theta_{\Gal}; t)=\int \frac{\dd\omega_{\obs}}{2\pi} e^{i\omega_{\obs} t}\tilde{h}_A^{[\Gal, \obs]}(x^{\mu}(\lambda), \theta_{\Gal}; \omega_{\obs})\,.
\ee
Similarly we introduce the Fourier transform of the GW signal emitted by a galaxy
\be\label{Proserpina2}
h_A^{[\Gal]}(x^{\mu}(\lambda), \theta_{\Gal}; t_{\Gal})=\int \frac{d\omega_{\Gal}}{2\pi} e^{i\omega_{\Gal} t_{\Gal}}\tilde{h}_A^{[\Gal]}(x^{\mu}(\lambda), \theta_{\Gal}; \omega_{\Gal})\,,
\ee
where $t_{\Gal}$ is the time measured in the galaxy frame.

We replace Eq. (\ref{Proserpina}) in the time average of Eq. (\ref{Ettore}). If the typical frequencies in the spectrum are such that $\omega_{\obs} \gg 1/T_{\obs}$, the time integral in the average amounts to a Fourier transform which brings a $\delta(\omega_{\obs}-\omega'_{\obs})$.  
We get
\begin{align}\label{Ettore3}
&\left[\dot{h}_A^{[\Gal, \obs]}(x^{\mu}(\lambda), \theta_{\Gal}; t)\dot{h}_A^{[\Gal, \obs]}(x^{\mu}(\lambda), \theta_{\Gal}; t)\right]\nn\\
&=8\pi^2 \int_0^{\infty} d\nu_{\obs}\nu_{\obs}^2 \frac{\Big| \tilde{h}_A^{[\Gal, \obs]}(x^{\mu}(\lambda), \theta_{\Gal}; \omega_{\obs})\Big|^2}{T_{\obs}}\,,
\end{align}
where we have used $\omega_{\obs}=2\pi \nu_{\obs}$. Using that $\nu_{\Gal}=\nu_{\obs}(1+z_{\Gal})$ and the reciprocity relation (\ref{reciprocity}), from Eq. (\ref{oem}), it follows that the Fourier components at observation and emission are related by  (omitting $\theta_{\Gal}$ in the notation) 
\begin{equation}\label{ringo}
\Big| \tilde{h}_A^{[\Gal, \obs]}(x^{\mu}(\lambda); \nu_{\obs})\Big|^2=\left(\frac{D_L^{\rm prox}}{D_A}\right)^2 \Big| \tilde{h}_A^{[\Gal]}(x^{\mu}(\lambda); \nu_{\Gal})\Big|^2\,.
\end{equation}
Hence (omitting $\theta_{\Gal}$ in the notation) we find
\begin{align}\label{Ettore3bis}
&\left[\dot{h}_A^{[\Gal, \obs]}(x^{\mu}(\lambda); t)\dot{h}_A^{[\Gal, \obs]}(x^{\mu}(\lambda); t)\right]\\
&=8\pi^2 \left(\frac{D_L^{\rm prox}}{D_A}\right)^2 \int_0^{\infty} \frac{d\nu_{\obs}\nu_{\Gal}^2}{[1+z_{\Gal}(\lambda)]^3} \frac{\Big| \tilde{h}_A^{[\Gal]}(x^{\mu}(\lambda); \omega_{\Gal})\Big|^2}{T_{\Gal}}\,,\nn
\end{align}
where  $T_{\Gal}=(1+z_{\Gal})^{-1} T_{\obs}$.  If the galaxy considered has only one gravitational-wave event,  $1/T_{\Gal}$ is the rate of events in the galaxy frame ${\cal R}_{\Gal}$. Hence, when considering all incoherent signals from individual events inside a galaxy, in analogy with what done in section II B of Ref.~\cite{Cusin:2017fwz},  we define
\begin{align}
&{\cal P}^{\Gal}_A (x^{\mu}(\lambda), \theta_{\Gal}; \omega_{\Gal})\nn\\
&\equiv\frac{\pi}{2G} {\cal R}_{[\Gal]} \Big| \tilde{h}_A^{[\Gal]}(x^{\mu}(\lambda), \theta_{\Gal}; \omega_{\Gal})\Big|^2\,,\nn\\
&\equiv\frac{\pi}{2G}\sum_i {\cal R}_{[\Gal,\,i]} \Big| \tilde{h}_A^{[\Gal,\,i]}(x^{\mu}(\lambda), \theta_{\Gal}; \omega_{\Gal})\Big|^2\,,
\end{align}
where $ {\cal R}_{[\Gal,i]}$ is the rate of gravitational-wave events of type $i$ (with $\tilde{h}_A^{[\Gal,i]}$ its associated strain) inside the galaxy considered.
With this last definition, substituting Eq. (\ref{Ettore3bis}) in Eq. (\ref{Ettore}), and writing the number of sources in an infinitesimal physical volume by using Eq. (\ref{NG}) and (\ref{Volume}), we find
\begin{widetext}
\be\label{Ettore5}
\frac{d^3\rho_{GW}}{d\nu_{\obs} d^2\bee_{\obs}}(t_{\obs}, \bx_{\obs}, \bee_{\obs})=\frac{c^2}{4\pi} \int d\lambda \int d\theta_{\Gal}  \frac{\sqrt{p_{\mu}(\lambda) p^{\mu}(\lambda)}}{\left[1+z_{\Gal}(\lambda)\right]^3}\,n_{\Gal}\left[x^{\mu}(\lambda), \theta_{\Gal}\right]  4\pi (D_L^{\rm prox})^2 \nu_{\Gal}^2 \sum_A {\cal P}_A^{\Gal}(x^{\mu}(\lambda), \theta_{\Gal}; \nu_{\Gal})\,.
\ee
\end{widetext}

Comparing this result with the expression of the energy density obtained in Ref. \cite{Cusin:2017fwz} using a different approach, Eq. (\ref{pippo2}), we conclude that the galaxy luminosity (per unit emitted frequency) is the related to the emitted strain by
\be\label{finale} 
\mathcal{L}_{\Gal}=4\pi (D_L^{\rm prox})^2 \nu_{\Gal}^2\sum_A  {\cal P}_A^{\Gal}(x^{\mu}(\lambda), \theta_{\Gal}; \nu_{\Gal})\,,
\ee
where we have used natural units (with $c=1$). In this form it is clear that $\nu_{\Gal}^2\sum_A  {\cal P}_A^{\Gal}$ is the energy flux  (per units of emitted frequency) close to the source, and physically it contains the incoherent superposition of the energy fluxes of the various sources inside the galaxy considered. Note that Eq.~(\ref{finale}) could have been obtained by a direct calculation using Eq.~(\ref{flux1}) and choosing a surface around the source, close enough to the source\footnote{We observe that (\ref{flux1}) is valid in the opposite regime with respect to the one we are considering here, i.e. for large distances from the source and using it close to the source is an extrapolation justified as long as strong gravity effects are negligible.}, and using
\begin{eqnarray}\label{Ettore3bis22}
&&\frac{1}{16 \pi G}\sum_A\left[\dot{h}_A^{[\Gal]}(x^{\mu}(\lambda); t)\dot{h}_A^{[\Gal]}(x^{\mu}(\lambda); t)\right]\nn\\
&&=\int d \nu_{\Gal }\nu_{\Gal}^2\sum_A  {\cal P}_A^{\Gal}(x^{\mu}(\lambda); \nu_{\Gal})\,,
\end{eqnarray}
hence showing the consistency of our derivation. 

We have proposed a new geometrical derivation of the energy density of the GW background. The final result is given by Eq. (\ref{Ettore5}), it is valid in a generic space-time geometry and it has been found starting from the definition of the GW background energy density, expressed as a time-average of the square of the received strain in a given direction, Eq. (\ref{flux33infinitesimal}), writing the total signal from a given direction as the sum of contributions of single sources, Eq. (\ref{master}) and finally expressing the signal received from a given source as a function of the emitted one. This last step is crucial and it has been fully detailed in sections \ref{eikonal} and \ref{sight}. By comparing the master expression of this paper, Eq. (\ref{Ettore5}), with our previous result (\ref{pippo2}), we found the expected expression~(\ref{finale}) for the luminosity of a given galaxy as a function of the emitted GW signal.

We conclude this section observing that the energy density of the AGWB, Eq.~(\ref{Ettore5}),  is a stochastic quantity since it depends on galaxy density, redshift and spatial displacement which are stochastic quantities. It can therefore be characterized in terms of its two-point correlation function introduced in Eq.~(\ref{Corr}). This correlation function is non vanishing due to the non vanishing correlator of the cosmological quantities (velocity, density and gravitational potential fields) which appear in Eq.~(\ref{Ettore5}) when specialized to a perturbed Friedmann-Lema\^{\i}tre  universe. The analytic expression of this correlation function in a universe with structures has been computed in our previous work Ref.~\cite{Cusin:2017fwz}.   
%
%
We have refined the simple model presented in section \ref{heuristic} and  our results confirmed what announced in Eqs.~(\ref{Icangivestupidnamesaswell1}) and (\ref{Corr}), following an heuristic argument. The anisotropies of the AGWB are characterized in terms of the energy density of the background and its two-point correlation function while the total amplitude of the GW signal received from different directions is uncorrelated. 
%

\section{Conclusions}\label{conclusion}

This article has clarified the relation between the GW signal that can be measured by interferometers and PTA and the energy density of the stochastic GW background. This provides a new and independent derivation of the result we established in Ref.~\cite{Cusin:2017fwz}. It explicitly shows why the AGWB energy density in a given direction enjoys angular correlations despite the fact that individual GW sources are incoherent and thus uncorrelated. The physical quantity that needs to be used to characterize AGWB anisotropies is not the GW signal, but the energy density in a given direction and its correlation function. The correlation arises from the cosmological variables and in particular the galaxy number density, the gravitational potentials and the cosmic velocity fields. All these fields inherit their stochasticity from the quantum initial conditions during inflation. This has two consequences: (1) the AGWB is correlated with other cosmological probes, such as weak lensing or galaxy number counts and (2) it encodes information on both cosmology and astrophysics (star formation rates, rates of binary mergers etc.; see e.g. Refs.~\cite{2017arXiv170909197D,Dvorkin:2016okx,Dvorkin:2016wac}). The computation of the angular power spectrum and numerical predictions is presented in Ref.~\cite{Cusin:2018rsq}.

A related question concerns the observability of this AGWB and the strategy to be developed so as to detect it. As already mentioned in the introduction of this paper, angular searches are implemented for both ground-based interferometers and pulsar-time array. In order to map the intensity of the AGWB,  techniques similar to those employed in radio astronomy for intensity mapping, {\it GW radiometry}~\cite{Mitra:2007mc}, can be used.

The recent detection by the Advanced Laser Interferometric Gravitational-wave Observatory (LIGO) of the gravitational wave sources GW150914~\cite{Abbott:2016blz} followed by GW151226 \cite{Abbott:2016nmj}  and GW170104 \cite{Abbott:2017vtc} and by the very recent observation of a black hole merging from both the LIGO and Virgo detectors \cite{Abbott:2017oio},  have pointed out that the rate and mass of coalescing binary black holes appear to be greater than many previous expectations. As a result, the stochastic background from unresolved compact binary coalescences is expected to be particularly loud. As explained in  Ref.~\cite{TheLIGOScientific:2016dpb},  the contribution of the AGWB coming from BBH binary systems has a high chance to be detected before Advanced LIGO will reach its final sensitivity.

%



\section*{Acknowledgements} 

We thank Joseph Romano for asking the question which led to this investigation and for interesting discussions during an early stage of this work. We thank Tania Regimbau, Joseph Romano, Alberto Sesana and Bernard Whiting for stimulating discussions and questions on our previous article~\cite{Cusin:2017fwz}. We thank Irina Dvorkin and Elisabeth 
Vangioni  for many discussions concerning the astrophysics of GW production and Camille Bonvin and Pierre Fleury for discussions about gravitational lensing. GC thanks the Institut d'Astrophysique de Paris for hospitality. This work has been done within the Labex ILP (reference ANR-10-LABX-63), part of the Idex SUPER, and received financial state aid managed by the Agence Nationale de la Recherche, as part of the programme Investissements d'avenir under the reference ANR-11-IDEX-0004-02. We acknowledge the financial support from the EMERGENCE 2016 project, Sorbonne Universit\'es, convention no. SU-16- R-EMR-61 (MODOG). The work of GC is supported by the Swiss National Science Foundation.

%
\bibliographystyle{utcaps}
\bibliography{BH+BH-refs.bib}
\end{document}